\documentclass[11pt]{article}

\usepackage{amssymb, amsmath, amscd}
\usepackage{graphicx}
\usepackage[latin1]{inputenc}
\usepackage{epsfig}
\usepackage{color}
\usepackage{subfig, color}
\usepackage{float}

\newtheorem{theorem}{Theorem}
\newtheorem{proposition}{Proposition}

\numberwithin{equation}{section}
\newtheorem{example}{Example}
\usepackage{tikz-cd}
\usepackage{rotating}

\begin{document}


\newcommand{\calA}{{\cal A}}
\newcommand{\calB}{{\cal B}}
\newcommand{\calF}{{\cal F}}
\newcommand{\calG}{{\cal G}}
\newcommand{\calR}{{\cal R}}
\newcommand{\calZ}{{\cal Z}}
\def\D{\mathcal{D}}
\def\L{\mathcal{L}}
\def\S{\mathcal{S}}
\def\I{\mathcal{I}}
\def\V{\mathcal{V}}
\def\E{\mathcal{E}}
\def\M{\mathcal{M}}

\def\A{\mathscr{A}}
\def\F{\mathscr{F}}
\def\G{\mathscr{G}}

\newcommand{\N}{{\mathbb N}}
\newcommand{\Z}{{\mathbb Z}}
\newcommand{\Q}{{\mathbb Q}}
\newcommand{\R}{{\mathbb R}}
\newcommand{\CC}{{\mathbb C}}

\newcommand{\K}{{\mathbb K}}
\newcommand{\kk}{{\mathrm k}}

\def\J{\mathrm{J}}
\def\catC{{\bf \mathrm{C}}}
\def\x{\mathrm{x}}
\def\a{\mathrm{a}}
\def\d{\mathrm{d}}
\def\Bi{\mathrm{Bi}}
\def\op{\mathrm{op}}
\def\res{\mathrm{res}}
\def\span{\mathrm{span}}

\newcommand{\C}{{\bf C}}
\newcommand{\Objects}{{\bf Objects}}
\newcommand{\Arrows}{{\bf Arrows}}
\newcommand{\Sets}{{\bf Sets}}

\def\2F1{\mbox{ $_2${F}$_1$}}
\def\1F1{\mbox{ $_1${F}$_1$}}
\def\1F2{\mbox{ $_1${F}$_2$}}
\def\0F1{\mbox{ $_0${F}$_1$}}

\def\GL{\mathrm{GL}}
\def\det{\mathrm{det}}
\def\SL{\mathrm{SL}}
\def\PSL{\mathrm{PSL}}
\def\PGL{\mathrm{PGL}}
\def\O{\mathrm{O}}

\def\gl{\mathfrak{gl}}
\def\g{\mathfrak{g}}
\def\h{\mathfrak{h}}
\def\frakM{\mathfrak{M}}

\newcommand{\Frac}[2]{\displaystyle \frac{#1}{#2}}
\newcommand{\Sum}[2]{\displaystyle{\sum_{#1}^{#2}}}
\newcommand{\Prod}[2]{\displaystyle{\prod_{#1}^{#2}}}
\newcommand{\Int}[2]{\displaystyle{\int_{#1}^{#2}}}
\newcommand{\Lim}[1]{\displaystyle{\lim_{#1}\ }}

\newenvironment{menumerate}{%
    \renewcommand{\theenumi}{\roman{enumi}}%
    \renewcommand{\labelenumi}{\rm(\theenumi)}%
    \begin{enumerate}} {\end{enumerate}}

\newenvironment{system}[1][]%
	{\begin{eqnarray} #1 \left\{ \begin{array}{lll}}%
	{\end{array} \right. \end{eqnarray}}

\newenvironment{meqnarray}%
	{\begin{eqnarray}  \begin{array}{rcl}}%
	{\end{array}  \end{eqnarray}}

\newenvironment{marray}%
	{\\ \begin{tabular}{ll}}
	{\end{tabular}\\}

\newenvironment{program}[1]%
	{\begin{center} \hrulefill \quad {\sf #1} \quad \hrulefill \\[8pt]
		\begin{minipage}{0.90\linewidth}}
	{\end{minipage} \end{center} \hrule \vspace{2pt} \hrule}

\newcommand{\entrylabel}[1]{\mbox{\textsf{#1:}}\hfil}
\newenvironment{entry}
   {\begin{list}{}%
   	{\renewcommand{\makelabel}{\entrylabel}%
   	  \setlength{\labelwidth}{40pt}%
   	  \setlength{\leftmargin}{\labelwidth + \labelsep}%
   	}%
   }%
   {\end{list}}

\newenvironment{remark}{\par \noindent {\bf Remark. }}
			{\hfill $\blacksquare$ \par}

\newenvironment{Pmatrix}
        {$ \left( \!\! \begin{array}{rr} }
        {\end{array} \!\! \right) $}

\newcommand{\fleche}[1]{\stackrel{#1}\longrightarrow}
\def\ssi{si et seulement si\ }
\newcommand{\tab}{\hspace*{\fill}}
\newcommand{\bs}{{\backslash}}
\newcommand{\eps}{{\varepsilon}}
\newcommand{\into}{{\;\rightarrow\;}}
\newcommand{\PD}[2]{\frac{\partial #1}{\partial #2}}
\def\Hat{\widehat}
\def\Bar{\overline}
\def\vect{\vec}
\def\fbar{{\bar f}}
\def\xbar{{\bar \x}}
\newcommand{\afaire}[1]{$$\vdots$$ \begin{center} {\sc #1} \end{center} $$\vdots$$ }
\newcommand{\pref}[1]{(\ref{#1})}

\def\Maple{{\sc Maple}}
\def\RG{{\sc Rosenfeld-Gr\"obner}}



\newcommand{\algf}{\sffamily}
\newcommand{\BEGIN}{{\algf begin}}
\newcommand{\END}{{\algf end}}
\newcommand{\IF}{{\algf if}}
\newcommand{\THEN}{{\algf then}}
\newcommand{\ELSE}{{\algf else}}
\newcommand{\ELIF}{{\algf elif}}
\newcommand{\FI}{{\algf fi}}
\newcommand{\WHILE}{{\algf while}}
\newcommand{\FOR}{{\algf for}}
\newcommand{\DO}{{\algf do}}
\newcommand{\OD}{{\algf od}}
\newcommand{\RETURN}{{\algf return}}
\newcommand{\PROCEDURE}{{\algf procedure}}
\newcommand{\FUNCTION}{{\algf function}}
\newcommand{\INDENTER}{{\algf si} \=\+\kill}

\newcommand{\target}{\mathop{\mathrm{t}}}
\newcommand{\source}{\mathop{\mathrm{s}}}
\newcommand{\trdeg}{\mathop{\mathrm{tr~deg}}}
\newcommand{\jet}[2]{\jmath_{#1}^{#2}}
\newcommand{\rank}{\operatorname{rank}}
\newcommand{\sign}{\operatorname{sign}}
\newcommand{\ord}{\operatorname{ord}}
\newcommand{\aut}{\operatorname{aut}}
\newcommand{\Hom}{\operatorname{Hom}}
\newcommand{\myhom}{\operatorname{hom}}
\newcommand{\codim}{\operatorname{codim}}
\newcommand{\coker}{\operatorname{coker}}
\newcommand{\rp}{\operatorname{rp}}
\newcommand{\leader}{\operatorname{ld}}
\newcommand{\card}{\operatorname{card}}
\newcommand{\Fr}{\operatorname{Frac}}
\newcommand{\RF}{\operatorname{\mathsf{reduced\_form}}}
\newcommand{\rang}{\operatorname{rang}}

\def \Id{\mathrm{Id}}

\def \diff{\mathrm{Diff}^{\mathrm{loc}} }
\def \diffg{\mathrm{Diff} }
\def \Esc{\mathrm{Esc}}

\newcommand{\initial}{\mathop{\mathsf{init}}}
\newcommand{\separant}{\mathop{\mathsf{sep}}}
\newcommand{\quo}{\mathop{\mathsf{quo}}}
\newcommand{\pquo}{\mathop{\mathsf{pquo}}}
\newcommand{\lcoeff}{\mathop{\mathsf{lcoeff}}}
\newcommand{\mvar}{\mathop{\mathsf{mvar}}}

\newcommand{\prem}{\mathop{\mathsf{prem}}}
\newcommand{\remp}{\mathrel{\mathsf{partial\_rem}}}
\newcommand{\remf}{\mathrel{\mathsf{full\_rem}}}
\renewcommand{\gcd}{\mathop{\mathrm{gcd}}}
\newcommand{\pairs}{\mathop{\mathrm{pairs}}}
\newcommand{\dd}{\mathrm{d}}
\newcommand{\ideal}[1]{(#1)}
\newcommand{\cont}{\mathop{\mathrm{cont}}}
\newcommand{\pp}{\mathop{\mathrm{pp}}}
\newcommand{\pgcd}{\mathop{\mathrm{pgcd}}}
\newcommand{\ppmc}{\mathop{\mathrm{ppcm}}}
\newcommand{\init}{\mathop{\mathrm{initial}}}

\bibliographystyle{amsalpha}

{

\title{
Minimizing polynomial functions on quantum computers
 }
\author{Raouf Dridi\footnote{rdridi@andrew.cmu.edu}, \, Hedayat Alghassi\footnote{halghassi@cmu.edu} ,\, Sridhar Tayur\footnote{stayur@cmu.edu} \\
\small  Quantum Computing Group\\
\small Tepper School of Business, Carnegie Mellon University, Pittsburgh, PA 15213\\}

\date{March 14, 2019}
\maketitle

\begin{abstract}
    This expository paper reviews some of the recent uses of computational algebraic geometry in classical and quantum optimization. The paper assumes an elementary background in algebraic geometry and adiabatic quantum computing (AQC), and concentrates on presenting concrete examples (with Python codes tested on a quantum computer) of applying algebraic geometry constructs: solving binary optimization, factoring, and compiling. Reversing the direction, we also briefly describe a novel use of quantum computers to compute Groebner bases for toric ideals. We also show how Groebner bases play a role in studying AQC at a fundamental level within a Morse theory framework. We close by placing our work in perspective, by situating this leg of the journey, as part of a  marvelous intellectual expedition that began with our ancients over 4000 years ago.
\end{abstract}

\tableofcontents

\section{Introduction}
The present paper tells the new story of the growing romance between  two protagonists: algebraic geometry \cite{cox} and adiabatic quantum computations \cite{Farhi472, vazirani}. An algebraic geometer, who has been introduced to the notion of Ising Hamiltonians \cite{Suzuki2013}, will quickly recognize  the attraction in this relationship. However, for many physicists, this connection could be surprising, primarily because of their  pre-conception that algebraic geometry is just a very abstract branch of pure mathematics. Although this is somewhat true--that is, algebraic geometry today studies variety of
sophisticated objects such as schemes and stacks--at heart, those are tools for studying the same problem that our ancients grappled with: solving systems of polynomial equations.

~~\\
A more known relationship is the one between
algebraic geometry and classical polynomial optimization which  dates back to the early  90s, with
the work of 
 B. Sturmfels and collaborators \cite{MR1363949, realalggeo}. The application of algebraic geometry to integer programming can be found  in \cite{Conti:1991:BAI:646027.676734, DBLP:journals/mp/TayurTN95,  DBLP:journals/mp/SturmfelsT97,  doi:10.1287/mnsc.46.7.999.12033}.  We take this occasion of an invited paper  to introduce both classical and quantum optimization applications of algebraic geometry (the latter, conceived by the authors)  through a number of concrete examples, with minimum possible abstraction, with the hope that it will serve as a teaser to join us in this leg of a marvelous expedition that began with the pioneering contributions of the Egyptian, Vedic, and   pre-Socrates Greek priesthoods.

\section{The profound interplay between algebra and geometry}
In mathematics, there are number of {\it dualities} that  differentiate it from other sciences. Through these dualities, data transcend   abstraction, allowing  different interpretations and access to different probing approaches. 
One of these is the duality between the {\it category} of (affine) algebraic varieties (i.e., zero loci of
systems of polynomial equations) and the category of (finitely generated with no nilpotent elements) commutative rings: 

\begin{equation}
	{\bf \left\{ \mbox{{\bf affine  algebraic  varieties}}\right\}} \simeq  \left\{ \mbox{{\bf coordinate  rings}}\right\}^{\mathrm{op}}
\end{equation}
Because of this equivalence,  we can go back and forth between the two equivalent descriptions, taking advantage of both worlds.

\begin{example}
Before we go any deeper, here is an example of an algebraic variety  
\begin{equation}
    \mathcal V := \mbox{ the unit circle in $\mathbb R^2$}.
\end{equation} The very same data (set of points at equal distance from the origin) is captured algebraically with the  coordinate ring  
\begin{equation}
  \mathbb Q[x, y]/ \langle x^2+y^2-1\rangle = \mbox{polynomials, in } x \mbox{ and } y, mod\,  (x^2+y^2-1).  
\end{equation}
As its name indicates, the coordinate ring   provides a coordinate system for the geometrical object $\mathcal{V}$. 
\end{example}

~~\\ 
We write $\mathbb Q[x_0, \ldots, x_{n-1}]$
for the ring of polynomials in $x_0, \ldots, x_{n-1}$ with rational coefficients (at some places, including the equivalence above, the field of coefficients $\mathbb Q$ should be replaced by its algebraic closure! In practice, this distinction is not problematic and can be safely swept under the rug). Let    
$\mathcal S$ be 
 a  set of polynomials $f\in \mathbb Q[x_0, \ldots, x_{n-1}]$. Let  $\mathcal V(S)$ denotes the  algebraic variety defined by 
the polynomials $f\in S$, that is, the set of common zeros of the equations $f=0, \, f\in \mathcal S$. The system $\mathcal S$ 
generates an {\it ideal} $\mathcal I $ by  taking all linear combinations over $\mathbb Q[x_0, \ldots, x_{n-1}]$ of all polynomials in  $\mathcal S$; we have $\mathcal V(\mathcal S)=\mathcal V(\mathcal I).$ The ideal $\mathcal I$ reveals the hidden polynomials that are the consequence of the generating  polynomials in $\mathcal S$. For instance, if one of the hidden polynomials is the constant polynomial 1 (i.e., $1\in \mathcal I$), then the system $\mathcal S$ is inconsistent (because $1\neq 0$).  To be precise, the set of all hidden polynomials is given
by the so-called {\it radical ideal} $\sqrt{\mathcal I}$, which is defined by \mbox{$\sqrt{\mathcal I} = \{g \in\mathbb Q[x_1, \ldots, x_n] |\,  \exists r\in \mathbb N: \, g^r\in \mathcal I \}$}. We have:
\begin{proposition}
   $\mathcal I(\mathcal V(\mathcal I))=\sqrt{\mathcal I}$.
 \end{proposition}
Of course, the radical ideal $\sqrt{\mathcal I}$ is infinite. Luckily, thanks to a prominent technical result (i.e., {\it Dickson's lemma}), it has a finite generating set i.e., a  
 {\it  Groebner basis} $\mathcal B$, which one might take to be a triangularization of the ideal $\sqrt{\mathcal I}$.  In fact, the computation of   Groebner bases  generalizes Gaussian elimination in linear systems. 
 \begin{proposition}
  $\mathcal V(\mathcal S)=\mathcal V(\mathcal I)=\mathcal V(\sqrt{\mathcal I})=\mathcal V(\mathcal B)$.
 \end{proposition} 
Instead of giving the technical definition of what a Groebner basis is (which can be found in \cite{cox} and in many other text books)   let us give an example (for simplicity, we use the term  ``Groebner bases" to refer to  {\it reduced} Groebner bases, which is,  technically what we are working with):
\begin{example}
 Consider the system by 
 $$
 	\mathcal S = \{x^2+y^2+z^2-4, x^2+2y^2-5, xz-1\}. 
 $$
 We want to solve $\mathcal S$. One way to do so is to compute a Groebner basis for~$\mathcal S$.
 In Figure \ref{code1}, the output of the cell number 4 gives a Groebner basis of $\mathcal S$. We can see that the initial system has been triangulized:  The last equation contains only the variable $z$, whilst the second has an additional variable, and so on. The variable $z$ is said to be eliminated with respect to the rest of the  variables. When computing the Groebner basis,  the underlying algorithm (Buchberger's algorithm)  uses the ordering $x>y>z$ (called lexigraphical ordering) for the computing of two internal calculations: cross-multiplications and Euclidean divisions.    The program tries to isolate $z$ first, then $z$ and $y$, and finally $x, y$, and $z$~(all variables). It is clear that different orderings yield different Groebner bases.

	 \begin{figure}[H]
    	\centering
    		\subfloat  
    		{{\includegraphics[height=8cm, width=16cm]{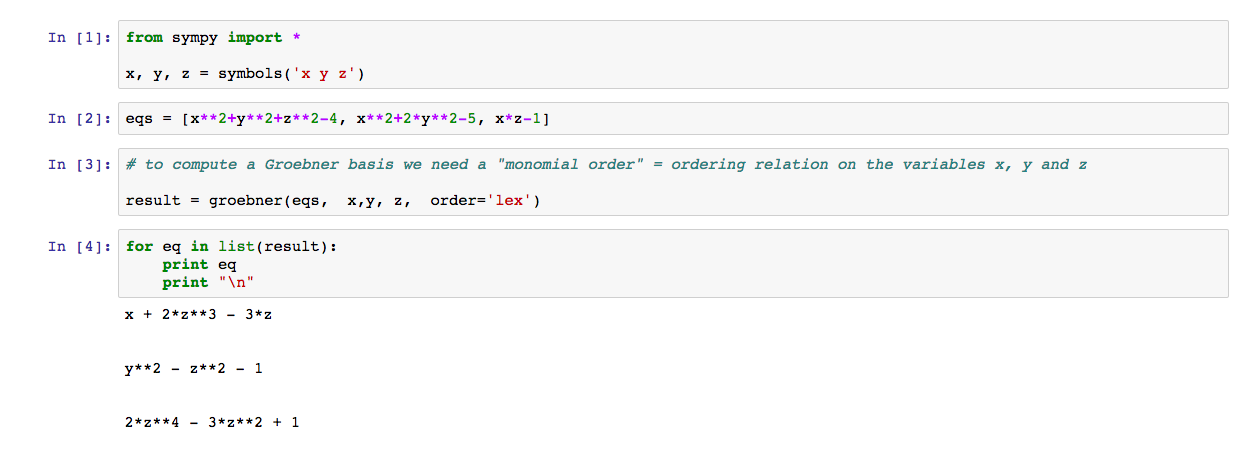} }}%
    		\caption{{Jupyter notebook for computing Groebner bases using   Python package {\sf sympy}. More efficient algorithms exist (e.g., \cite{Faugere:2002:NEA:780506.780516, Faugere199961}).  }}
           \label{code1}
 	 \end{figure}%
 	 
\end{example}
 
~~\\
The mathematical power of Groebner bases doesn't stop at solving systems of algebraic equations. The applicability of Groebner bases  goes well beyond this:   it gives necessary and sufficient conditions for the existence of solutions. Let us illustrate  this with an example.

\begin{example}
Consider the following 0-1 feasibility problem   
\begin{system}
x_1 + x_1 +x_3 &=& b_1,\\
x_1 + a_2 x_2  &=& b_2,
\end{system}%
with $x_i \in \{0,1\}$ for $i=1,2,3.$ By putting the variables 
$a_2, b_1$, and $b_2$ to the rightmost of the ordering, we obtain the set of all $a_2, b_1$, and $b_2$ for which the system is feasible. The notebook in Figure \ref{feasible} shows the details of the calculations as well as the conditions on  the variables $a_2, b_1$, and~$b_2$.

	 \begin{figure}[H]
    	\centering
    		\subfloat  
    		{{\includegraphics[height=10cm, width=16cm]{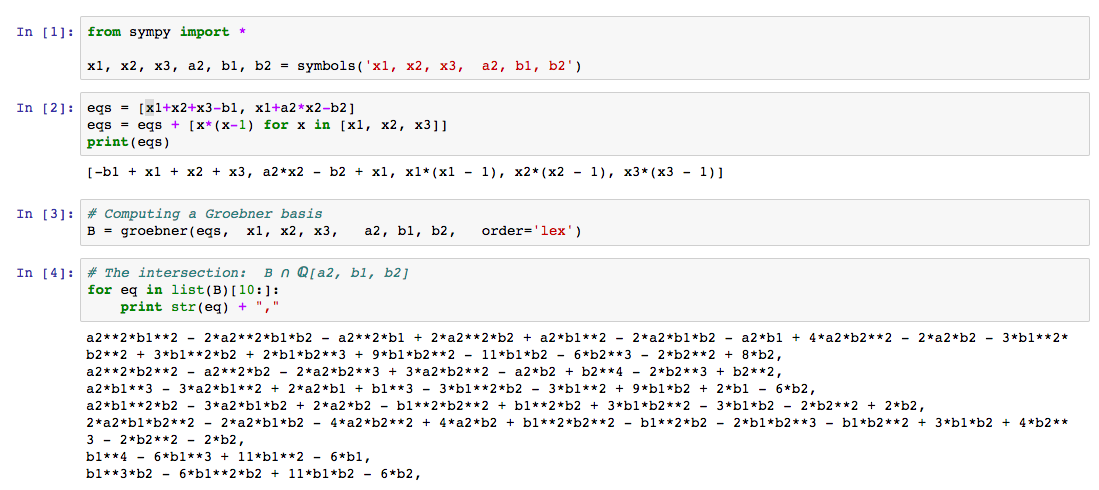} }}%
    		\caption{Necessary and sufficient conditions for existence of feasible solutions.}
  	  	\label{feasible}
 	 \end{figure}%
\end{example}

~~\\This machinery can be put in more precise wording as follows: 
\begin{theorem}
Let $\mathcal{I}\subset \mathbb Q[x_0, \ldots, x_{n-1}]$ be an ideal, and let $\mathcal{B}$ be a reduced Groebnber basis of~$\mathcal{I}$
with respect to the lex order $x_0\succ \ldots \succ x_{n-1}$. Then, for every $0\leq l\leq n-1$, the set 
\begin{equation}\label{intersectionB}
	\mathcal{B}\cap \mathbb Q[x_{l}, \ldots, x_{n-1}]
\end{equation}
is a  Groebner basis of the ideal $\mathcal{I}\cap Q[x_{l}, \ldots, x_{n-1}]$.
\end{theorem}
As previously mentioned,   this elimination theorem  is used to obtain
the complete set of conditions on the variables $x_{l}, \ldots, x_{n-1}$, such that the ideal $\mathcal{I}$ is not empty. For instance, if the ideal represents a system of algebraic equations and these equations are (algebraically) dependent on certain parameters, then the  intersection \pref{intersectionB}
gives {\it all} necessary and sufficient conditions for the existence of solutions.  
\section{The innate role of algebraic geometry in binary optimization}
By now, it should not be surprising to see algebraic geometry emerges when optimizing polynomial functions. Here, we expand on this with two examples of how algebraic geometry solves  the binary polynomial optimization
   \begin{equation}\label{P}
		(\mathcal P):\, argmin_{(y_0, \cdots, y_{m-1})\in  \{0,1\} ^m} \,  f(y_0, \cdots, y_{m-1}).
	\end{equation} 
The first method we review here was introduced in \cite{doi:10.1287/mnsc.46.7.999.12033} (different from another previous method that is studied in \cite{DBLP:journals/mp/TayurTN95}, which we discuss in a later section). The second method we review here is new, and is an adaptation of the method described in \cite{realalggeo} to the binary case.  

\subsection{A general  method for solving binary optimizations}
The key idea is to consider   the ideal
$$
  \mathcal I = \{z- f(y_0, \cdots, y_{m-1}), \, 
  y_0^2-y_0,\, \cdots,\,  y_{m-1}^2-y_{m-1}\},
  $$
where we note the appearance of the variable $z$. This new variable covers the range of the function $f$. Consequently, if we compute a Groebner basis with an elimination ordering in which $z$ appears at the rightmost, we obtain a polynomial in $z$ that gives all values of $f$. Take, then, the smallest of those values and substitute in the rest of the basis and solve.   
\begin{example}
Consider the following problem  
\begin{system}
   argmin_{y_i\in \{0, 1\}} &&  \, y_1 + 2y_2 + 3y_3 +3y_4,\\[3mm]
    && \,y_1 + y_2 + 2y_3 + y_4 =3. 
 \end{system} %
Figure \ref{code2} details   the solution. 
	 \begin{figure}[H]
    	\centering
    		\subfloat  
    		{{\includegraphics[height=10cm, width=16cm]{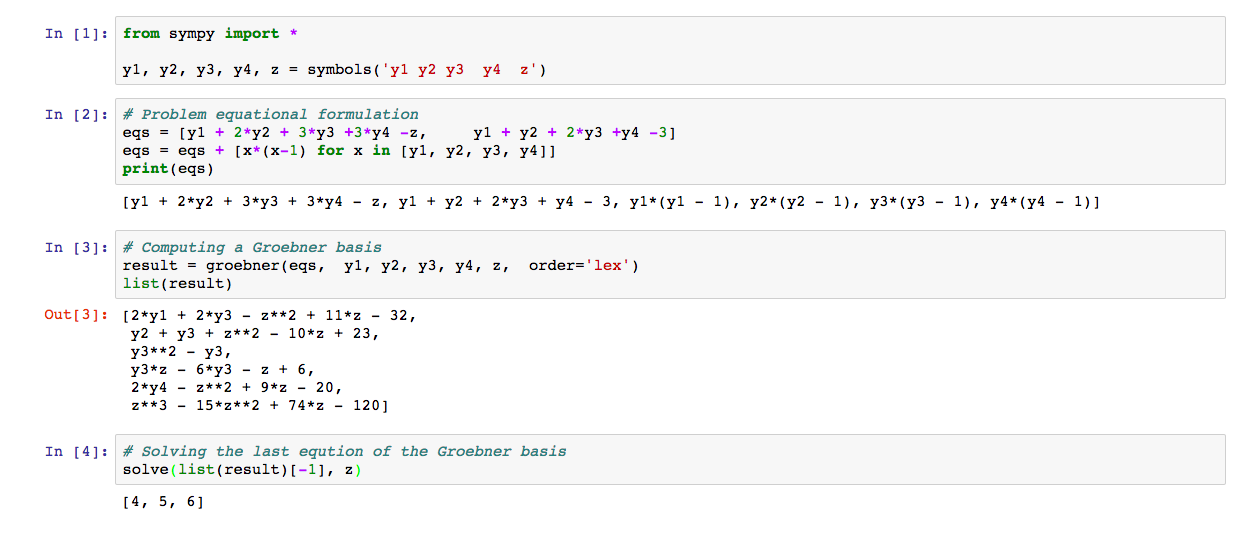} }}%
    		\caption{Solving optimization problems with Groebner bases. Although, the cost function  is linear here,
    		the method works for any polynomial function.}
  	  	\label{code2}
 	 \end{figure}%
\end{example}

\subsection{A second general method for solving binary optimizations}
An important construction that comes with the  cost function $f$ is the gradient ideal.  This is a valuable   additional information that we will use in the resolution of the problem $(\mathcal P)$. Now, because the arguments of the cost function~$f$ are binary, we need to make sense of the derivation of the function~$f$. This is taken care of by the introduction of the function  
\begin{equation}
    \tilde f :=   f + \sum_{i=0}^{m-1} \alpha_i^2 y_i(y_i-1),
\end{equation}
where are real numbers with  $|\alpha_i|>1$. We can now go ahead and define {\it the gradient ideal} of $f$ as
\begin{equation}
    \tilde {\mathcal I} := <\partial_{y_0} \tilde f, \cdots, \partial_{\alpha_{m-1}^2}\tilde f >.
\end{equation}
The variety $\mathcal{V}(\tilde {\mathcal I})$ gives the set of local minima of the function $f$. 
Its coordinate ring is the  {\it residue algebra} 
\begin{equation}
    A:=\mathbb Q[y_0,\ldots, y_{m-1},\alpha_0,\ldots, \alpha_{m-1}]/\tilde{\mathcal   I}.
\end{equation}
Let us define 
  the linear map
  \begin{eqnarray}
m_{\tilde{f}}:& A& \rightarrow A\\\nonumber
							& g& \mapsto \tilde{f}  g	
\end{eqnarray}
Because the number of local minima is finite, the residue algebra 
$A$ is finite-dimensional. Because of this, the following is true \cite{cox}:  
\begin{itemize}
	\item {The values of $\tilde f $, on the set of critical points $\mathcal V(\tilde{\mathcal{I}})$,
	are given by the  eigenvalues of the matrix $m_{\tilde{f}}$.}
	\item The eigenvalues of $m_{y_i}$ and $m_{\alpha_i}$ give the coordinates of the points of  $\mathcal V(\tilde{\mathcal{I}})$. 
	\item If $v$ is an eigenvector for $m_{\tilde f}$, then it is also an eigenvector for $m_{y_i}$ and $m_{\alpha_i}$ for $0\leq i\leq m-1$. 
\end{itemize}

~~\\
We need to compute a basis for $A$. This is done by first computing a Groebner basis for $\tilde{\mathcal I}$ and then extracting the standard monomials (i.e., the monomials in 
$\mathbb Q[y_0,\ldots, y_{m-1},\alpha_0,\ldots, \alpha_{m-1}]$ that are not divisible by the leading term of any
element in the Groebner basis).  In the simple example below, we do not need to compute any Groebner basis, because
 $\tilde{\mathcal I}$ is a Groebner basis with respect to $plex (\alpha, y)$. 
\begin{example}
    We illustrate this on 
$$
	f = 2+7\,x_{{4}}+2\,x_{{3}}+2\,x_{{4}}x_{{3}}-2\,x_{{3}}x_{{2}}-x_{{1}}-4
\,x_{{4}}x_{{1}}-2\,x_{{3}}x_{{1}}+x_{{2}}x_{{1}},
$$
where $x_i \in \{0, 1\}$. 
A basis for the residue algebra $A$ is 
given by the set of the 16 monomials 
$$
\{ 1,x_{{4}},x_{{3}},x_{{4}}x_{{3}},x_{{2}},x_{{4}}x_{{2}},x_{{3}}x_{{2}
},x_{{3}}x_{{2}}x_{{4}},x_{{1}},x_{{4}}x_{{1}},x_{{3}}x_{{1}},x_{{1}}x
_{{3}}x_{{4}},x_{{2}}x_{{1}},x_{{4}}x_{{1}}x_{{2}},x_{{1}}x_{{3}}x_{{2
}},x_{{1}}x_{{3}}x_{{2}}x_{{4}}\}.
$$
The matrix $m_{\tilde{f}}$ is  

$$
m_{\tilde{f}} := 
\left( \begin {array}{cccccccccccccccc}
2&7&2&2&0&0&-2&0&-1&-4&-2&0&1
&0&0&0\\ \noalign{\medskip}0&9&0&4&0&0&0&-2&0&-5&0&-2&0&1&0&0
\\ \noalign{\medskip}0&0&4&9&0&0&-2&0&0&0&-3&-4&0&0&1&0
\\ \noalign{\medskip}0&0&0&13&0&0&0&-2&0&0&0&-7&0&0&0&1
\\ \noalign{\medskip}0&0&0&0&2&7&0&2&0&0&0&0&0&-4&-2&0
\\ \noalign{\medskip}0&0&0&0&0&9&0&2&0&0&0&0&0&-4&0&-2
\\ \noalign{\medskip}0&0&0&0&0&0&2&9&0&0&0&0&0&0&-2&-4
\\ \noalign{\medskip}0&0&0&0&0&0&0&11&0&0&0&0&0&0&0&-6
\\ \noalign{\medskip}0&0&0&0&0&0&0&0&1&3&0&2&1&0&-2&0
\\ \noalign{\medskip}0&0&0&0&0&0&0&0&0&4&0&2&0&1&0&-2
\\ \noalign{\medskip}0&0&0&0&0&0&0&0&0&0&1&5&0&0&-1&0
\\ \noalign{\medskip}0&0&0&0&0&0&0&0&0&0&0&6&0&0&0&-1
\\ \noalign{\medskip}0&0&0&0&0&0&0&0&0&0&0&0&2&3&-2&2
\\ \noalign{\medskip}0&0&0&0&0&0&0&0&0&0&0&0&0&5&0&0
\\ \noalign{\medskip}0&0&0&0&0&0&0&0&0&0&0&0&0&0&0&5
\\ \noalign{\medskip}0&0&0&0&0&0&0&0&0&0&0&0&0&0&0&5\end {array}
 \right)
$$
We obtain the following eigenvalues for $m_{\tilde{f}}$:
$$
\{0, 1, 2, 4, 5, 6, 9, 11, 13\}.
$$	 
This is also the set of values that $f$ takes on $\mathcal V(\tilde{\mathcal I)}$. The eigenvector $v$ that corresponds
to the eigenvalue 0 is the column vector
$$
	v := \left(1, 0, 1, 0, 1, 0, 1, 0, 1, 0, 1, 0, 1, 0, 1, 0 \right)^T.
$$
 This eigenvector is used to find the coordinates of $\hat x\in\mathcal V(\tilde{\mathcal I})$ that  minimize~$f$. The coordinates of the global minimum $\hat x = (\hat x_0, \ldots, \hat x_{m-1})$  are defined by $m_{x_i} v=\hat x_i v$, and this gives $x_1=x_2=x_3=1, \, x_4=0$, and ${\alpha_1= 2\alpha_2 =\alpha_3=2, \, \alpha_4=5}$.

\end{example}

\section{Factoring on quantum annealers} 
This section reviews  the use of the Groebner bases machinery in the factoring problem on current quantum annealers (introduced in    \cite{raouffactorization}). We need to deal with 
three key constraints: first, the number of available qubits. Second, the limited dynamic range for the allowed values of
the couplers (i.e., coefficients of the quadratic monomials in the cost function),  and third, the sparsity of the hardware graph. 

\subsection{Reduction}
In general, reducing a polynomial function $f$
into a quadratic function necessitates the injection of extra  variables (the minimum reduction is given in terms of toric ideals \cite{cmu1}). 
However, in certain cases, the reduction to QUBOs can be done without the additional variables. This is the example of the Hamiltonian that results from the long multiplication \cite{raouffactorization}.  
In fact, in addition to reduction, we can also adjust the coefficients to be within the dynamic range needed, at the same time. 
	Consider the quadratic polynomial
	$$
		H_{ij} := Q_{{i}}P_{{j}}+S_{{i,j}}+Z_{{i,j}}-S_{{i+1,j-1}}-2\,Z_{{i,j+1}},
	$$ 
	with the binary variables $P_{{j}},Q_{{i}},S_{{i,j}},S_{{i+1,j-1}},Z_{{i,j}},Z_{{i,j+1}}$. The goal is to solve $H_{ij}$ (obtain its zeros) by converting it into a QUBO. Instead of directly squaring the function $H_{ij}$ (naive approach) and then reducing the cubic function result into a quadratic function by adding extra variables, we  compute a Groebner basis $\mathcal B$ of the system
		$$
			\mathcal S = \{H_{ij}\} \cup \{{x^2-x, \, x\in \{P_{{j}},Q_{{i}},S_{{i,j}},S_{{i+1,j-1}},Z_{{i,j}},Z_{{i,j+1}}}\}\},
		$$
		and look for a positive quadratic polynomial ${H_{ij} }^+= \sum_{t\in \mathcal B|\, deg(t)\leq 2} a_t t$ in the ideal generated by $\mathcal{S}$. Note that
		 global minima of ${H_{ij} }^+$ are the zeros of $H_{ij}$. 
The Groebner basis $\mathcal B$ is
{
\begin{eqnarray}\nonumber
 t_1 &:= &Q_{{i}}P_{{j}}+S_{{i,j}}+Z_{{i,j}}-S_{{i+1,j-1}}-2\,Z_{{i,j+1}}\label{HijInB},\\[2mm]\nonumber
 t_2 &:= &\left( -Z_{{i,j+1}}+Z_{{i,j}} \right) S_{{i+1,j-1}}+ \left( Z_{{i,j+1
}}-1 \right) Z_{{i,j}},\\[2mm]\nonumber
 t_3 &:= & \left( -Z_{{i,j+1}}+Z_{{i,j}} \right) S_{{i,j}}+Z_{{i,j+1}}-Z_{{i,j+1
}}Z_{{i,j}},\\[2mm]\nonumber
 t_4 &:= &\left( S_{{i+1,j-1}}+Z_{{i,j+1}}-1 \right) S_{{i,j}}-S_{{i+1,j-1}}Z_{
{i,j+1}},\\[2mm]\nonumber
 t_5 &:= &\left( -S_{{i+1,j-1}}-2\,Z_{{i,j+1}}+Z_{{i,j}}+S_{{i,j}} \right) Q_{{
i}}-S_{{i,j}}-Z_{{i,j}}+S_{{i+1,j-1}}+2\,Z_{{i,j+1}},\\[2mm]\nonumber
 t_6 &:= &\left( -S_{{i+1,j-1}}-2\,Z_{{i,j+1}}+Z_{{i,j}}+S_{{i,j}} \right) P_{{
j}}-S_{{i,j}}-Z_{{i,j}}+S_{{i+1,j-1}}+2\,Z_{{i,j+1}},\\[2mm]\nonumber
&& \mbox { in addition to 3 more cubic polynomials.}
\end{eqnarray}}%
We take 
 $
 	{H_{ij} }^+ = \sum_{t\in \mathcal B|\, deg(t)\leq 2} a_t t,
 $
and solve for the $a_t$. 
 We can   require that the coefficients  $a_t$ are subject to the dynamic range allowed by the quantum processor 
(e.g.,  the absolute values of the coefficients of  ${H_{ij}^{+}}$,  with respect to the variables  $P_{{j}},Q_{{i}},S_{{i,j}},S_{{i+1,j-1}},Z_{{i,j}}$, and $Z_{{i,j+1}}$,
be within $[1-\epsilon, \, 1+\epsilon]$).  The ensemble of these constraints translates into a simple real optimization problem for the coefficients $a_t$.   
 
\subsection{Embedding}
The connectivity graph of the resulting quadratic polynomial $H^{+}_{ij}$ is   the complete graph $K_6$. Although embedding this into current  architectures  is not evident, the situation becomes better with upcoming architectures~{(e.g.,  D-Wave's next generation quantum processors \cite{pegasus})}.

\section{Compiling on quantum annealers}
Compiling the problem $(\mathcal P)$  in AQC, consists of two steps: reduction of the problem's polynomial function into a quadratic function (covered above) and later embedding the graph of the quadratic function inside the quantum annealer's hardware graph. This process  can be fully automatized using the language of algebraic geometry \cite{cmu1}. We review here the key points of this automatization, through a simple example. 

~~\\
Let us first explain what is meant by  embeddings (and introduce the subtleties that come with).    Consider the following optimization problem that we wish to solve on the D-Wave 2000Q quantum processor: 
	\begin{equation}
		(\mathcal P_\star):\, argmin_{(y_0, \cdots, y_{m-1})\in  \{0,1\} ^m} \quad  y_0 \sum_{i=1}^8 c_i y_i. 
	\end{equation}
	 \begin{figure}[H]
    	\centering
    		\subfloat  
    		{{\includegraphics[width=3cm]{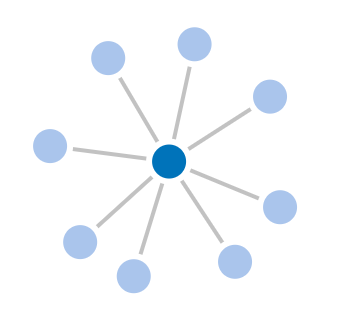} }}%
    		\qquad 
    		\subfloat  
    	{{\includegraphics[width=4cm]{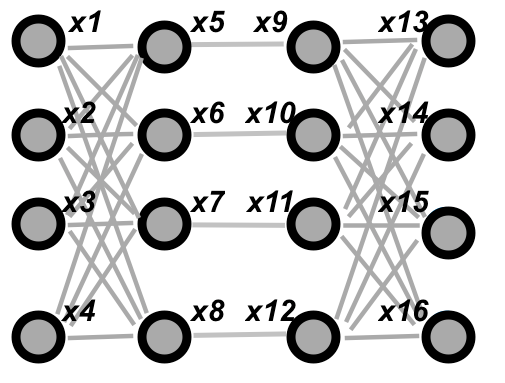} }}%
    		\, 
    		\subfloat  
    	{{\includegraphics[width=4cm]{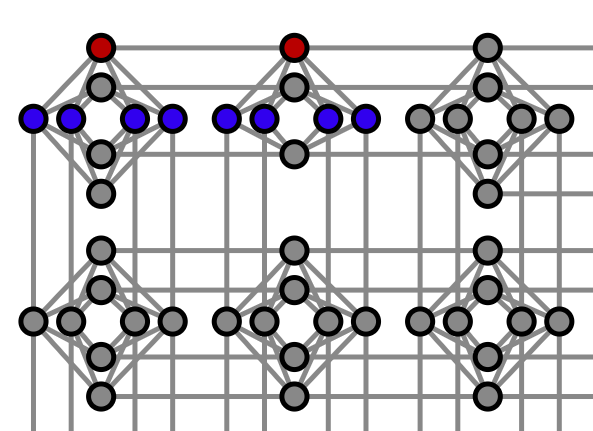} }}%
    		\caption{{\small (Left) The logical graph of the objective function in $(\mathcal P_\star)$, can not be embedded inside Chimera graph. (Center) We blow up the central node into edges $(x_5, x_9)$ and redistribute the surrounding nodes. (Right) Embedding inside an actual D-Wave 2000Q quantum annealer; in   red, the chain of qubits representing the logical qubit $y_0$. The missing qubits  are faulty.  
    		}}
  	  	\label{star}
 	 \end{figure}%
~~\\
Before we start annealing, we need to map the logical variables $y_i$ to the physical qubits of the hardware. Similarly, the quadratic term $c_i y_0y_i$ needs to be mapped into a coupling between physical qubits with strength given by the coefficient $c_i$.    Not surprisingly,  this mapping can not always be  a simple matching--because of the sparsity of the hardware graph (Chimera in our case).  This is true for  our simple example; 
the degree of the central node is 8, so a direct matching inside Chimera, where the maximum degree is 6, is not feasible. Thus,  we stretch the definition of embedding. Instead, we allow nodes to elongate or, as an algebraic geometer will say, to  {\it blow up}. In particular, if we blow up the central node $y_0$ into an edge, say the edge $(x_5, x_9)$, we can then  redistribute the surrounding nodes $y_1,\cdots, y_8$,  at these two duplicates of $y_0$.   In general, one needs a sequence of blow ups, which turns out to be a hard problem.  What makes the problem even harder is that not all embeddings are equally valued.  It is important to choose embeddings that have, among others, smaller chains, as illustrated in  Figure~\ref{chainBreaking}. Of course, this is in addition to minimizing the overall number of physical qubits used. 
 
   \begin{figure}[H]
    \centering
    \subfloat  
    {{\includegraphics[width=6cm]{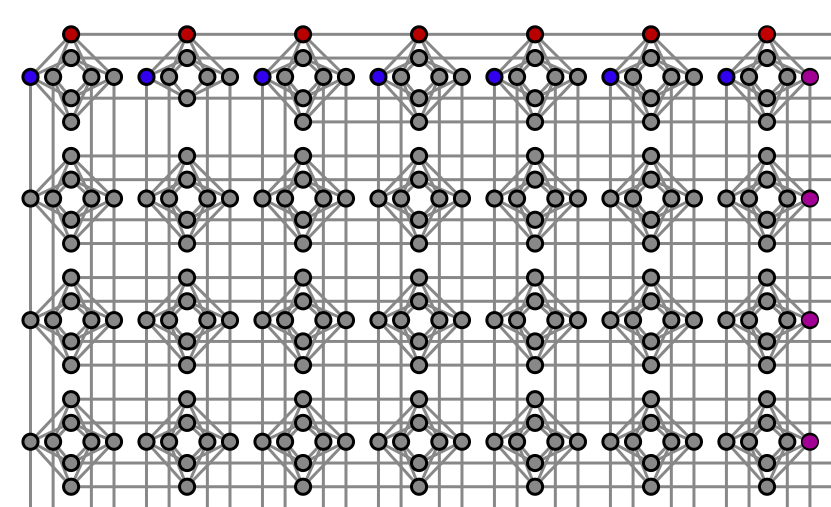} }}%
    \caption{{\small The depicted  embedding (for  the problem $(\mathcal P)_*$) has two long chains that don't persist through the adiabatic evolution (in D-Wave 2000Q processor). In this case, the quantum processor fails to return an answer.   }}
    \label{chainBreaking}
  \end{figure}
  
\subsection {Embeddings as fiber-bundles}
One way to think about embedding the logical graph $Y$ into the hardware graph $X$ is in terms of {\it fiber-bundles}. This {\it equational} formulation makes the connection with algebraic geometry. 
The general form of such  fiber-bundles  is

\begin{eqnarray}\label{rep2}
	\pi(x_i)   &=& \sum _{ij} \alpha_{ij} y_j  \\\nonumber
	&\mbox {with }& \sum _{ij} \alpha_{ij} = \beta_i, \quad
	  \alpha_{ij_1} \alpha_{ij_2} = 0,\quad
	  \alpha_{ij} (\alpha_{ij}-1)=0,
\end{eqnarray}
where the binary number $\beta_i$ is  1 if  the physical qubits $x_i$ is used  and 0 otherwise. We write
 $domain(\pi)= {\bf Vertices}(X)$ and $support(\pi)={\bf Vertices}(X^\beta)$ with $X^\beta\subset_{subgraph} X$. The fiber  of the map $\pi$ at $y_j\in {\bf Vertices}(Y)$ is   given by 
  \begin{equation}
  	\pi^{-1}(y_j) = \phi(y_j) = \{ x_i\in {\bf Vertices}(X)|\quad \alpha_{ij}=1 \}.
  \end{equation}
 The conditions on  
  the parameters $\alpha_{ij}$ guarantee that   fibers don't intersect (i.e., $\pi$ is well defined map). In addition to these conditions, two more conditions need to be satisfied: (i)  {\sf Pullback Condition}: the logical graph $Y$ embeds entirely inside $X$ (ii) {\sf Connected Fiber Condition}:  each fiber is a connected subgraph (of $X$). We will not go into the details of these conditions, which can be found in \cite{cmu1}. We illustrate this in a simple example.

\begin{example}
	Consider the two graphs in Figure \ref{simpleExple}.  In this case, equations (\ref{rep2})
	are given by
  	\begin{eqnarray}
	 && \alpha_{{1,1}}\alpha_{{1,2}}, \, \alpha_{{1,1}}\alpha_{{1,3}}, \, 
\alpha_{{1,2}}\alpha_{{1,3}},
\\
&&
\alpha_{{2,1}}\alpha_{{2,2}},\, \alpha_{{2,1}}\alpha_{{2,3}},\, \alpha_{{2,2}}\alpha_{{2,3}},
\\
&&
\alpha_{{3,1}}\alpha_{{3,2}},\, \alpha_{{3,1}}\alpha_{{3,3}},\, \alpha_{{3,2}}\alpha_{{3,3}},
\\
&&
\alpha_
{{4,1}}\alpha_{{4,2}},\, 
\alpha_{{4,1}}\alpha_{{4,3}},\, \alpha_{{4,2}}
\alpha_{{4,3}},
\\
&&
\alpha_{{5,1}}\alpha_{{5,2}},\, \alpha_{{5,1}}\alpha_{{5,3}},\, \alpha_{{5,2}}\alpha_{{5,3}}, 
	\end{eqnarray}
and 
\begin{eqnarray}\nonumber
&&
\alpha_{{1,1}}+\alpha_{{1,2}}+\alpha_{{1,3}}-\beta_{{1}},
\quad 
\alpha_{{2,1}}+\alpha_{{2,2}}+\alpha_{{2,3}}-\beta_{{2}},
\quad 
\alpha_{{3,1}
}+\alpha_{{3,2}}+\alpha_{{3,3}}-\beta_{{3}},
\\\nonumber
&&
\alpha_{{4,1}}+\alpha_{{4,
2}}+\alpha_{{4,3}}-\beta_{{4}}, 
\quad
\alpha_{{5,1}}+\alpha_{{5,2}}+\alpha_{{
5,3}}-\beta_{{5}}. 
\end{eqnarray}
  \begin{figure}[H] %
    \centering
    \subfloat 
    {{\includegraphics[width=4cm]{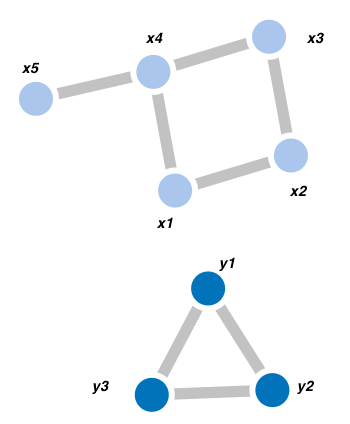} }}%
    \caption[]{{The set of  {\it all} fiber bundles $\pi:X\rightarrow Y$
    defines an algebraic variety. This variety is given by the Groebner basis \pref{GBSimpleExple}.
	}}%
    \label{simpleExple}%
\end{figure}
~~\\
The {\sf Pullback Condition} reads
{\small 
\begin{eqnarray}\nonumber
	&&-1+\alpha_{{4,1}}\alpha_{{5,2}}+\alpha_{{3,1}}\alpha_{{4,2}}+
\alpha_{{1,1}}\alpha_{{2,2}}+\alpha_{{3,2}}\alpha_{{4,1}}+\alpha_{{1,2
}}\alpha_{{2,1}}+\alpha_{{1,2}}\alpha_{{4,1}}+\alpha_{{2,2}}\alpha_{{3
,1}}+\alpha_{{1,1}}\alpha_{{4,2}}+\alpha_{{2,1}}\alpha_{{3,2}}+\alpha_
{{4,2}}\alpha_{{5,1}},
\\\nonumber
&&
-1 +\alpha_{{3,3}}\alpha_{{4,1}}+\alpha_{{1,3}}
\alpha_{{2,1}}+\alpha_{{2,3}}\alpha_{{3,1}}+\alpha_{{4,1}}\alpha_{{5
,3}}+\alpha_{{1,3}}\alpha_{{4,1}}+\alpha_{{1,1}}\alpha_{{2,3}}+\alpha_
{{4,3}}\alpha_{{5,1}}+\alpha_{{2,1}}\alpha_{{3,3}}+\alpha_{{3,1}}
\alpha_{{4,3}}+\alpha_{{1,1}}\alpha_{{4,3}},
\\\nonumber
&& 
-1 +\alpha_{{3,3}}\alpha_{{4,2
}}+\alpha_{{1,2}}\alpha_{{2,3}}+\alpha_{{1,2}}\alpha_{{4,3}}+\alpha_{{
1,3}}\alpha_{{2,2}}+\alpha_{{1,3}}\alpha_{{4,2}}+\alpha_{{2,3}}\alpha_
{{3,2}}+\alpha_{{2,2}}\alpha_{{3,3}}+\alpha_{{4,2}}\alpha_{{5,3}}+
\alpha_{{3,2}}\alpha_{{4,3}}+\alpha_{{4,3}}\alpha_{{5,2}}.  \nonumber
\end{eqnarray}
}
Finally, the {\sf Connected Fiber Condition} is given by
{\small
\begin{eqnarray}\nonumber
&&
-\alpha_{{1,1}}\alpha_{{2,1}}\alpha_{{5,1
}},-\alpha_{{1,1}}\alpha_{{3,1}}\alpha_{{5,1}},-\alpha_{{1,2}}\alpha_{
{2,2}}\alpha_{{5,2}},
-\alpha_{{1,2}}\alpha_{{3,2}}\alpha_{{5,2}},-
\alpha_{{1,3}}\alpha_{{2,3}}\alpha_{{5,3}},-\alpha_{{1,3}}\alpha_{{3,3
}}\alpha_{{5,3}}
\\\nonumber
&&
-\alpha_{{2,1}}\alpha_{{3,1}}\alpha_{{5,1}},-\alpha_{
{2,1}}\alpha_{{4,1}}\alpha_{{5,1}},-\alpha_{{2,2}}\alpha_{{3,2}}\alpha
_{{5,2}},-\alpha_{{2,2}}\alpha_{{4,2}}\alpha_{{5,2}},-\alpha_{{2,3}}
\alpha_{{3,3}}\alpha_{{5,3}},-\alpha_{{2,3}}\alpha_{{4,3}}\alpha_{{5,3
}},
\\\nonumber
&&
\quad
\alpha_{{2,1}}\alpha_{{5,1}},\alpha_{{2,2}}\alpha_{{5,2}},
\alpha_{{2,3}}\alpha_{{5,3}}.
\end{eqnarray}	
}%
We can then use the elimination theorem to obtain all embeddings of $Y$ inside $X$ (by putting the variables $\beta_i$ to the right most of the elimination order). A part of the r  Groebner basis  is given by 
\begin{eqnarray}\label{GBSimpleExple} 
\mathcal B &=&   \left\{ \beta_{{1}}-1,\beta_{{2}}-1,\beta_{{3}}-1,\beta_{{4}}-1, 
 \beta^2_5-\beta_5,\, \alpha_{ij}^2 -\alpha_{ij}, \,\right.
\\[3mm]\nonumber
&&
\alpha_{{1,2}}\alpha_{{1,3}},\alpha_{{1,2}}\alpha_{{3,2}},
\alpha_{{1,3}}\alpha_{{3,3}},\alpha_{{2,2}}\alpha_{{2,3}},\alpha_{{2,2
}}\alpha_{{4,2}},\alpha_{{2,2}}\alpha_{{5,2}},\alpha_{{2,3}}\alpha_{{4
,3}},\alpha_{{2,3}}\alpha_{{5,3}},\alpha_{{3,2}}\alpha_{{3,3}}, \alpha_{{4,2}}\alpha_{{4,3}},
\\[3mm]\nonumber
&& 
\alpha_{{4,2}}\alpha_{{5,3}},
\alpha_{{4,3}}\alpha_{{5,2}},\alpha_{{5,2}}\alpha_{{5,3}},\alpha_{{4,2}}\alpha_{{5,2
}}-\alpha_{{5,2}},\alpha_{{4,2}}\beta_{{5}}-\alpha_{{5,2}},\alpha_{{4,
3}}\alpha_{{5,3}}-\alpha_{{5,3}}, 
\\\nonumber
&&
\quad \quad \vdots\\\nonumber
&&
-\alpha_{{2,2
}}\alpha_{{5,3}}-\alpha_{{3,2}}\alpha_{{5,3}}+\alpha_{{1,2}}\beta_{{5}
}+\alpha_{{2,2}}\beta_{{5}}+\alpha_{{3,2}}\beta_{{5}}+\alpha_{{3,3}}
\beta_{{5}}+\alpha_{{5,2}}+\alpha_{{5,3}}-\beta_{{5}}\left.\right \}. 
\end{eqnarray}
 In particular, the intersection $\mathcal B\cap \mathbb Q[\beta] =(\beta_{{1}}-1,\beta_{{2}}-1,\beta_{{3}}-1,\beta_{{4}}-1, {\beta_{{5}}}^{2}-\beta_{{5}})$ gives
the two $Y$ minors (i.e., subgraphs $X^\beta$) inside $X$.  
 The remainder of $\mathcal B$ gives
the explicit expressions of the corresponding mappings.
\end{example}

\subsection {Symmetry reduction}
Many of the embeddings aquired using the above method, are redundant. We can eliminate this redundancy in a mathematically elegant way  using the theory of invariants \cite{olver_1999} (on top of the algebraic geometrical formulation). First, we fold the hardware graph along its symmetries
and then proceed as before.  This amounts to re-expressing the quadratic form of the hardware graph in terms of the
invariants of the symmetry.

 \begin{example}
 Continuing with the same example: The  quadratic form of $X$ is:
	\begin{equation}
		Q_X(x) = x_{1}x_{2}+x_{2}x_{3}+x_{3}x_{4}+x_{1}x_{4}+x_{4}x_{5}. 
	\end{equation}
	Exchanging the two nodes $x_1$ and $x_3$ is a symmetry for $X,$ and the quantities $K=x_1 + x_3,  \, x_2, \,  x_4,$ and  $x_5$ are invariants of this symmetry.	
	In terms of these invariants, the quadratic function $Q_X(x)$,  takes the simplified form: 
	\begin{equation}
		Q_X(x, K) = K x_{2}+K x_{4} +x_{4}x_{5},  
	\end{equation}
	which shows (as expected) that graph $X$ can be folded into a chain (given by the new nodes $[x_2, K, x_4, x_5]$). 
	The surjective homomorphism $\pi:X\rightarrow Y$  now takes  the form
	\begin{eqnarray}
		K &=& \alpha_{01}y_1 +\alpha_{02}y_2 +\alpha_{03}y_3.\\
		x_i &=&  \alpha_{i1}y_1 +\alpha_{i2}y_2 +\alpha_{i3}y_3 \mbox { for } i=2, 4, 5. 
	\end{eqnarray}%
The table below compares the computations of the surjections $\pi$  with and without the use of invariants:
	 
	{\small 
	\begin{center}
	\begin{tabular}{l|l|l}
						  	 					& original coords 	&  invt coords\\\hline
		Time for computing a  Groebner basis (in secs)	& 0.122 			& 0.039\\
		Number of defining equations					& 58				& 30\\
		Maximum degree in the defining eqns  		&  3				& 2\\
		Number of variables	in the defining eqns 		& 20				&12\\
		Number of solutions							& 48				& 24\\
		\hline
 	\end{tabular}  
	\end{center}
	}
  	
~~\\
In particular, the number of solutions is down
 to 24, that is, four (non symmetric) minors times
 the six symmetries of the logical graph $Y$.
\end{example}

\section{Quantum computing for algebraic geometry}
Here we give an example that goes in the opposite direction of what we have covered so far. We show how quantum computers can be used to compute algebraic geometrical structures that are exponentially hard to   compute classically. Our attention   is directed to a prominent type of polynomial ideals; the so-called {\it toric ideals} and their Groebner bases.  In the context of the theory of integer optimization, this gives a novel quantum algorithm for solving IP problems (a quantum version of Conti and Traverso algorithm \cite{Conti:1991:BAI:646027.676734}, that is used in \cite{DBLP:journals/mp/TayurTN95}). As a matter of fact, the procedure  which we are about to describe can be used to construct the full {\it Groebner fan} \cite{MR1363949, cox} of a given toric ideal. We leave the technical details for a future work. A related notion is the so-called Graver basis which extends toric Groebner bases in the context of convex optimization. A hybrid classical-quantum algorithm for computing Graver bases is given in \cite{hedayat}. 

~~\\ Toric ideals are ideals 
 generated by differences of monomials. Because of this, their Groebner bases enjoy a clear structure  given by   kernels of integer matrices.  Specifically,  let $A =(a_1, \cdots, a_n)$ be any integer $m\times n$-matrix ($A$ is called configuration matrix). Each 
 column $ {{\bf a}}_i = (a_{1i}, \cdots, a_{ni})^T$ is identified with  a Laurent monomial $y^{ {{\bf a}}_i} = y_1^{a_{1i}}\cdots y_m^{a_{ni}}$. In this case, the toric ideal $\mathcal J_A$ associated   with the configuration $A$ is the kernel of the algebra homomorphism 
 \begin{eqnarray}
	\mathbb Q[x] \rightarrow \mathbb Q[y]\\
	 x_i \mapsto y^{ {{\bf a}}_i}.
 \end{eqnarray}
 From this it follows that the toric ideal  $\mathcal J_A$ is generated by the binomials $x^{{{\bf u}}_+} - x^{{{\bf u}}_-},$ where the vector 
  ${\bf u} = {{\bf u}}_+ - {{\bf u}}_{-}\in {\mathbb Z^+}^n\oplus {\mathbb Z^+}^n$
  runs over all integer vectors in  $\mathrm{Ker}_\mathbb Z A$, the  kernel of the matrix~$A$.	 
 It is not hard to see that the elimination theorem that we have used repeatedly can also be used here to compute a Groebner basis for 
 the toric ideal  $\mathcal J_A$. 
 
 ~~\\
 Now we  explain how AQC (or any quantum optimizer such as Quantum Approximate Optimization Algorithm, QAOA \cite{1411.4028}) can be used to compute  Groebner bases for the toric ideal  $\mathcal J_A$. 
 The example we choose is taken from \cite{cox}-Chapter 8. The matrix $A$ is given by  \begin{equation}
    A =
    \begin{pmatrix}
    4&5&1&0\\
    2&3&0&1
\end{pmatrix}
\end{equation}
The kernel is easily obtained (with polynomial complexity). It is the two  dimensional $\mathbb Z-$vector space spanned with
$
    \left( (1,0,-4,-2),\, (0,1,-5,-3)\right). 
$
We define $u=(a,b,-5\,b-4\,a,-3\,b-2\,a)$, which is a linear combination (over $\mathbb Z$) of the two vectors. As in \cite{cox}, we consider the lexographical ordering $plex( w_4, w_3, w_2, w_1)$  represented by the matrix order
 \begin{equation}
    M =
    \begin{pmatrix}
    0&0&0&1\\
    0&0&1&0\\
    0&1&0&0\\
    1&0&0&0
\end{pmatrix}. 
\end{equation}
The cost function is given by the square of the Euclidean norm of the vector
$M u^t$. Figure~\ref{toric} details the solution of this optimization problem on D-Wave 2000Q quantum processor. Each solution has twelve entries, and is of the form  
$[a_{0,+}, a_{0,-}, \cdots, b_{2,+}, b_{2,-}]$, corresponding to the binary decomposition of the integers $a=\sum_{i=0,1,2} (a_{i,+}   -   a_{i,-}) 2^i$ and ${b=\sum_{i=0,1,2} (b_{i,+}   -   b_{i,-}) 2^i}$. Qubits marked -1 are not used, so they should be considered equal to zero. The collection of all these solutions translates into the sought Groebner basis 
\begin{system}[{\mathcal B =}]
{{ w_2}}^{2}{{ w_3}}^{2}-{{ w_1}}^{3},\\
{{ w_1}}^{4}{ w_4}-{{ w_2}}^{3}{ w_3},\\
{ w_1}\,{ w_4}\,{ w_3}-{ w_2},\\
{ w_2}\,{ w_4}\,{{ w_3}}^{3}-{{ w_1}}^{2},\\
{{ w_3}}^{4}{{ w_4}}^{2}-{ w_1}. 
\end{system}

 \newpage
 
 \begin{figure}[h]
    	\centering
    		\subfloat  
    		{{\includegraphics[scale=0.55]{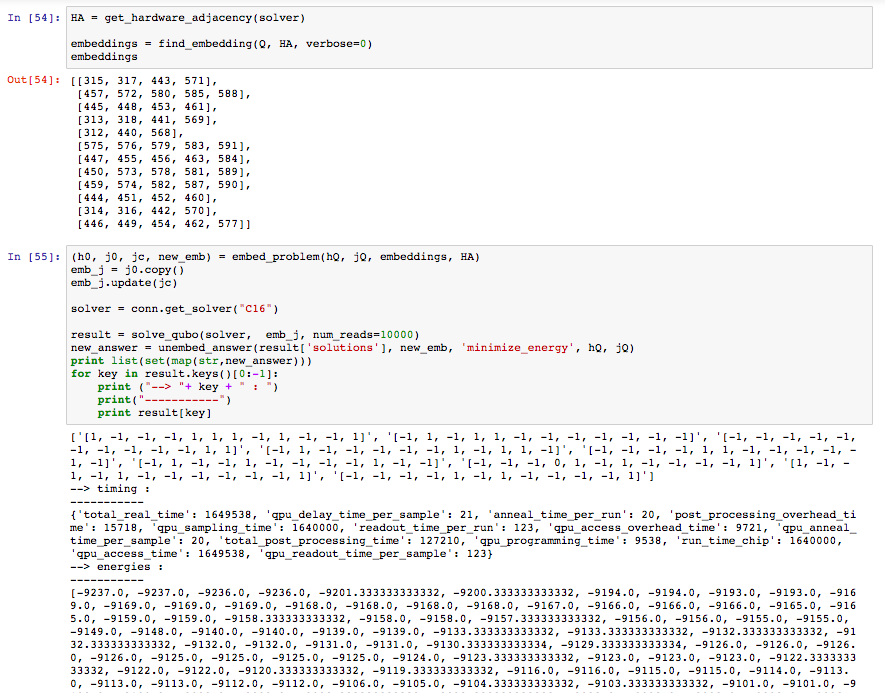}  }}%
    		\caption{Computation of toric Groebner bases on the D-Wave 2000Q quantum processor.}
  	  	\label{toric}
 	 \end{figure}%

 \newpage

\section{Groebner bases in the fundamental theory of AQC}
The role of the so-called {\it anti-crossings} \cite{Suzuki2013, vonNeumann1993} 
in AQC is well understood. This is expressed as the total adiabatic evolution time being inversely proportional to the square of the minimum energy difference between the two lowest energies of 
the given Hamiltonian.  This minimum is attained at anti-crossings. In this last section, we connect anti-crossings to the theory of Groebner bases (through a quick detour to {\it Morse theory} \cite{PMIHES_1988__68__99_0, witten1982}). 

~~\\
Consider the time dependant Hamiltonian: 
\begin{equation}\label{H}
    H(s) = (1-s) H_{initial}  + s H_{final},
\end{equation}
To the Hamiltonian (\ref{H}), we assign the function $f$ given by the characteristics polynomial: 
\begin{equation}\label{detH}
    f(s, \lambda) = det(H(s)-\lambda I), 
\end{equation}
where $I$ is the identity $n\times n-$matrix.  The important role that the function $f$ plays in AQC is described in \cite{cmu2, cmu3}. In particular, anti-crossings are now mapped into {\it saddle points} of the function $f$. This is the starting point of the connection with Morse theory, which is explored in details in \cite{cmu2, cmu3}.  
Here we explain how anti-crossings can be described using Groebner bases. The key fact is that
the function $f$ is a  polynomial function of $s$ and $\lambda$, and so is any partial derivative of $f$.  Recall that a critical point $p$ of $f$ is a point at which the differential map $df$ is the zero map--that is, the gradient of $f$ vanishes at $p$. A critical point is said to be non degenerate (e.g., a saddle point) if the determinant of the  Hessian of $f$ at $p$ is not zero. Define the ideal $\mathcal I$
generated by   the two polynomials $\partial_sf$ and $\partial_\lambda f$. It is clear that the variety of $\mathcal I$ gives the set of  all  critical points of $f$. To capture the non degeneracy, we need to {\it saturate} the ideal $\mathcal I$ with the polynomial
$det(\mathrm{Hessian}(f))$. This saturation is the ideal given by all polynomials in $\mathcal I$ that vanish for all the zeros of $\mathcal I$ that are not zeros of  $det(\mathrm{Hessian}(f))$. 
In other words, 
 a point $p$ is a non degenerate critical point of the function $f$ if and only if the remainder  ${\sf NormalForm}_\mathcal B (det(\mathrm{Hessian}(f)))$  is not zero at $p$, where $\mathcal B$
is a Groebner basis for the ideal $\mathcal I$.  

\section{Summary and discussion}
As we mentioned in the Introduction, we are travelers in a journey that our ancients started. Evidence of ``practical mathematics" during 2200 BCE in the Indus Valley has been unearthed that indicates proficiency in {\it geometry}. Similarly, in Egypt (around 2000 BCE) and Babylon (1900 BCE), there is good evidence (through the {\it Rhind Papyrus} and clay tablets, respectively) of capabilities in geometry and {\it algebra}. After the fall of the Indus Valley Civilization (around 1900 BCE), the Vedic period was especially fertile for mathematics, and around 600 BCE, there is evidence that {\it magnetism} (discovered near Varanasi) was already used for practical purposes  in medicine (like pulling arrows out of warriors injured in battle), as written in {\it Sushruta Brahmana}. Magnetism was also independently discovered by pre-Socrates Greeks, as evidenced by the writings of Thales (624-548 BCE), who, along with Pythagoras (570-495 BCE), was also quite competent in geometry. Indeed, well before Alexander (The Great), and the high points of Hellenistic Greek period), there is evidence that the Greeks were already doing some type of {\it algebraic geometry}.

~~\\
Algebra, which is derived from the Arabic word meaning completion or ``reunion of broken parts",
reached a new high watermark during the golden age of Islamic mathematics around 10th Century AD. For example, Omar Khayyam (of the {\it Rubaiyat} fame) solved cubic equations. The next significant leap in algebraic geometry, a {\it Renaissance}, in the 16th and 17th century, is quintessentially European: Cardano, Fontana, Pascal, Descartes, Fermat. The 19th and 20th Century welcomed the dazzling contributions of Laguerre, Cayley, Reimann, Hilbert, Macaulay, and the Italian school led by Castelnuov, del Pezzo, Enriques, Fano, and Severio. Modern algebraic geometry has been indelibly altered by van der Waerden, Zariski, Weil, and in 1950s and 1960s, by Serre and Grothendieck. Computational algebraic geometry begins with the Buchberger in 1965 who introduced Groebner bases (the first conference on computational algebraic geometry was in 1979).

~~\\
Magnetism simply could not be explained by classical physics, and had to wait for quantum mechanics. The workhorse to study it mathematically is the {\it Ising} model, conceived in 1925.
{\it Quantum computing} was first introduced by Feynman in 1981 \cite{Feynman1982}. The study of Ising models that formed a basis of physical realization of a quantum annealer (like D-Wave devices) can be traced to the 1989 paper by Ray, Chakrabarti and Chakrabarti \cite{PhysRevB.39.11828}. Building on various adiabatic theorems of the early quantum mechanics and complexity theory, adiabatic quantum computing was proposed by Farhi et al in 2001. 

~~\\
Which brings us to current times. The use of computational algebraic geometry (along with  Morse homology, Cerf theory and Gauss-Bonnet theorem from differential geometry) in the study of adiabatic quantum computing, and numerically testing our ideas on D-Wave quantum processors, which is a physical realization of an Ising model, is conceived by us, the authors, of this expository article. Let us close with the Roman poet {\it Ovid} (43 BC-17~AD): ``Let others praise ancient times; I am glad I was born in these."

~~\\


 \bibliography{c1}
\end{document}